\documentclass[reprint,amsmath,amssymb,aps,usenatbib,useAMS,nofootinbib,floatfix,prd]{revtex4-2}

\usepackage{graphicx}
\usepackage{dcolumn}
\usepackage{bm}
\usepackage{color}
\usepackage{lipsum}
\usepackage{multirow}
\usepackage{centernot}
\usepackage[dvipsnames,table]{xcolor}
\usepackage{hyperref}
\hypersetup{urlcolor=RoyalBlue,citecolor=ForestGreen,linkcolor=RedViolet,colorlinks=true}
\usepackage[normalem]{ulem}
\usepackage{aas_macros}
\usepackage{gensymb}
\usepackage{multirow}
\usepackage{hhline}
\usepackage{tabularray}

\newcommand{\eg}{\textit{e.g.}}

\definecolor{background}{RGB}{255,255,255} 
\definecolor{PineGreen}{RGB}{1, 121, 111} 

\begin{document}
\pagecolor{background}

\title{Skewness as a Probe of Gravity: Real and Redshift Space Counts-In-Cells}

\author{Pawe\l{} Drozda} \email{(pdrozda, hellwing, bilicki)@cft.edu.pl}
\author{Wojciech A.~Hellwing}
\author{Maciej Bilicki}

\affiliation{Center for Theoretical Physics, Polish Academy of Sciences, Al. Lotników 32/46,
02-668 Warsaw, Poland}

\date{\today}

\begin{abstract}
We study the counts-in-cells reduced skewness $s_3$ for dark matter, halo, and galaxy distributions in both real and redshift space, using the ELEPHANT (\textit{Extended LEnsing PHysics with ANalytical ray Tracing}) suite of $N$-body simulations. We compare General Relativity (GR) with two extended (EG) gravity models: $f(R)$ gravity with chameleon screening and the normal-branch Dvali–Gabadadze–Porrati (nDGP) model with Vainshtein screening. We quantify the suppression of $s_3$ by redshift-space distortions (RSD), finding that while small-scale skewness is strongly reduced, the $F5$ model retains a $\sim4\%$ deviation from GR in galaxy samples, corresponding to a $2\sigma$ significance. We show that the ratio $s_3^{\mathrm{RSD}}/s_3^{\mathrm{real}}$ is approximately independent of the gravity model across tracers and redshifts. Our results demonstrate that real-space predictions can help reliably infer redshift-space skewness in both GR and extended gravity, providing a new tool for testing gravity with current and forthcoming galaxy redshift surveys.
\end{abstract}


\maketitle

\section{Introduction}\label{sec.intro}

The large-scale structure of the Universe provides a rich testing ground for cosmological models, particularly through statistical analysis of density and velocity fields. Traditionally, two-point statistics such as the power spectrum and two-point correlation function have been instrumental in constraining cosmological parameters \citep[\eg][]{peebles1980the,Cole2005,Tegmark2006,SDSS2017,Beutler2011,DESI2025}. Yet, these measures alone cannot fully capture the complexity of the Cosmic Web. Higher-order statistics, such as skewness and kurtosis, complement two-point measures by probing the non-linear gravitational clustering process, which can significantly differ between General Relativity (GR) and various Extended Gravity (EG) models (in the literature also dubbed as  'Modified Gravity')\citep{Hellwing2013a, Hellwing2017PRD, Drozda2022}.

Recent studies highlight that these higher-order statistics are sensitive to subtle deviations from GR, making them powerful tools for testing structure formation scenarios in EG theories \citep{BoseTaruya2018, Hellwing2020the, Cataneo2022, Sugiyama2023}. Measurements of skewness and kurtosis in particular can reveal modifications to gravity or dark energy beyond the standard $\Lambda$CDM model. Despite their promise, such analyses have predominantly focused on theoretical configuration space\cite{Hellwing2010,Hellwing2013a,Hellwing2017PRD, Fazolo2022}. However, actual observations readily access only galaxy positions in redshift space, where peculiar velocities introduce distortions that complicate the interpretation of clustering statistics \citep{scoccimarro2004redshift, kaiser1987clustering, HernandezAguayo2019, Valogiannis2020}. These redshift-space distortions (RSD), 
while adding further complications to the galaxy clustering pattern, also carry valuable information about the growth rate of cosmic structure. Accurate modeling of RSD is thus crucial for interpreting observations and testing gravity theories \citep{KuruvillaPorciani2019, Kodwani2019}.

A systematic exploration of higher-order statistics in redshift space, particularly skewness, remains lacking but can become important for the next generation of robust cosmological tests. Constraints on cosmological parameters based on different statistics, scales, and phenomena are pivotal for internal consistency checks and for identifying potential unknown systematic effects that could limit the precision of these tests. Here, the potential of higher-order cumulants of the matter, halo, and galaxy density fields, remains largely unexploited and not fully tapped.

Addressing this gap, we perform a detailed study of skewness in both real and redshift space, comparing GR with two prominent EG scenarios: the $f(R)$ gravity model with chameleon screening and the normal-branch Dvali-Gabadadze-Porrati (nDGP) gravity model employing Vainshtein screening. We analyze how the transition from real to redshift space influences the detectability of gravity-induced differences, investigating separately the dark matter, halo, and galaxy distributions. Hence, this paper is a natural extension of our previous study focused on the angular clustering moments.\cite{Drozda2022} 

Recent simulation-based studies have demonstrated that modifications to gravity can significantly alter higher-order clustering statistics, such as skewness and kurtosis, in both dark matter and halo distributions \citep{Hellwing2013a, Hellwing2017PRD, arnalte2017real, Einasto2021}. In particular, deviations in the reduced skewness $s_3$ have been identified as a sensitive probe of screening mechanisms, with models like $f(R)$ and nDGP showing scale-dependent signatures in real-space clustering. However, translating these signatures into redshift space remains challenging. While RSD can suppress or mask EG signals at small scales \citep{HernandezAguayo2019, Valogiannis2020}, new approaches incorporating realistic velocity distributions \citep{KuruvillaPorciani2019} and angular clustering statistics \citep{Drozda2022} offer promising avenues for observational tests. Furthermore, recent theoretical developments emphasize the need to move beyond two-point functions, using bispectra and parity-breaking correlations \citep{BoseTaruya2018, Kodwani2019}, to robustly probe deviations from GR in upcoming surveys. Nonetheless, a comprehensive and systematic study of higher-order cumulants in redshift space across different tracer populations and gravity models remains largely unexplored. Our work aims to fill this gap.

This paper is structured as follows: Section \ref{sec.sim} describes the simulations and catalogs; Section \ref{sec.calc} introduces the statistical tools employed; Section \ref{sec.res} presents our analysis and findings, emphasizing skewness across real and redshift spaces; and Section \ref{sec.concl} discusses the broader implications of our results.

\section{Simulations and Catalogs}\label{sec.sim}
In this work, we use the \textsc{ELEPHANT} (\textit{Extended LEnsing PHysics with ANalytical ray Tracing}) $N$-body simulation suite \cite{cautun2018the}, performed with the ECOSMOG code \cite{li2012ecosmog}, a modified version of RAMSES adapted for EG models. 
The simulations evolve $1024^3$ dark matter pseudo-particles within a $(1024/h~\mathrm{Mpc})^3$ comoving box, generated from five independent realizations
of the initial power spectrum.
The assumed background cosmology corresponds to the WMAP9 results \cite{hinshaw2013nine}, with parameters 
$\Omega_M = 0.281$, $\Omega_{\Lambda} = 0.719$, and $h = 0.697$.
Dark matter halos were identified using the ROCKSTAR halo finder \cite{behroozi2013the}, which employs a phase-space friends-of-friends 
algorithm.
We also utilize \textsc{ELEPHANT}-dedicated galaxy mock catalogs, constructed using the Halo Occupation Distribution (HOD) method. 
The HOD parameters were calibrated to reproduce the projected two-point correlation function of BOSS CMASS DR9 galaxies 
\cite{alam2021towards} and to match the number density of a volume-limited sample \cite{white2011the}.
The mocks are constructed to recover the assumed satellite-to-centrals ratio (varying from $10-15\%$ across redshifts see \citet{hernandez2019large} fro details),
giving us a chance to study separately all-galaxy and centrals-only samples.

For biased tracers, we construct three halo samples, H1, H2, and H3, characterized by their mean number densities
of $10^{-3}$, $5\times 10^{-4}$, and $10^{-4}$ $(h/\mathrm{Mpc})^3$, respectively, following the definitions 
in \cite{garciafarieta2021probing}.
For each model, sample, and redshift, the desired number density is achieved by rejecting halos with masses smaller 
than an established threshold $M_{\mathrm{min}}$ that fulfills the target number density criterion. The exact 
mass cuts are listed in Table~\ref{tab.halomasscuts}, along with the corresponding number densities for both 
the halo and galaxy catalogs.
The mass cuts are based on the halo virial mass, $M_{200c}$, defined as the mass enclosed within the volume where 
the mean density exceeds 200 times the critical density.
For galaxies, the catalogs exhibit slight variations in number density; therefore, in Table~\ref{tab.halomasscuts}, 
we show the range of values rather than a single number.

\begin{table}[h]
\caption{Mass cuts in halo catalogs for a given sample, model and redshift.
The column corresponding to object number densities shows values both for
halos (H1-H3 samples) and galaxies.}

\begingroup
\setlength{\tabcolsep}{1pt}
\renewcommand{\arraystretch}{1.3}

\begin{tabular}{c c c c c c c c}
\hline
\multirow{2}{*}{\textbf{$z$}}&\multirow{2}{*}{\textbf{Sample}}&\textbf{$n$}&&
\multicolumn{3}{c}{\textbf{$M_{min} [10^{13} M_{\odot}/h]$}}&\\
&&[$10^{-4}$ $(\mathrm{Mpc}/h)^{-3}$]&GR&F6&F5&N5&N1\\
\hline\hline
\multirow{4}{*}{}&H1&10&0.26& 0.30& 0.30&  0.27& 0.27\\
{0}&H2&5&0.58& 0.64& 0.66& 0.58& 0.59\\
{}&H3&1&2.73& 2.80& 3.22& 2.76&  2.84\\
{}&Galaxies&3.8-3.9&-&-&-&-&-\\
\hline
\multirow{4}{*}{}&H1&10&0.26& 0.30& 0.31& 0.26& 0.27\\
{0.3}&H2&5&0.56& 0.59& 0.64& 0.56& 0.58\\
{}&H3&1&2.41& 2.44&  2.83&  2.43& 2.53\\

{}&Galaxies&3.1-3.3&-&-&-&-&-\\
\hline
\multirow{4}{*}{}&H1&10&0.23& 0.26& 0.30& 0.23& 0.25\\
{0.5}&H2&5&0.53& 0.55& 0.62& 0.54& 0.55\\
{}&H3&1&2.13& 2.14& 2.49& 2.15& 2.25\\

{}&Galaxies&3.1-3.3&-&-&-&-&-\\
\hline
\end{tabular}

\endgroup
\label{tab.halomasscuts}
\end{table}

\subsection{Extended Gravity}
In addition to the standard $\Lambda$CDM cosmology with GR-based structure formation, we consider 
two popular phenomenological extensions of General Relativity, implemented within the \textsc{ELEPHANT} suite. These are:

\begin{itemize}
    \item The $f(R)$ gravity model with the chameleon screening mechanism \cite{khoury2004chameleon,brax2008f(R)},
    \item The normal branch of Dvali-Gabadadze-Porrati (nDGP) gravity with the Vainshtein screening mechanism \cite{vainshtein1972problem,babichev2013introduction}.
\end{itemize}

These two EG models provide a testbed for cosmological inference based on real and redshift-space reduced cumulants, and we will compare their predictions with those of GR.

\subsection{Redshift-Space}
In this study, we work both in configuration (real) space and in redshift space. The latter can be constructed using 
information about the positions and radial velocities of objects relative to the observer. To construct our redshift-space 
catalogs, we employ the distant observer approximation, that is, the galaxy/halo line-of-sight (LOS) directions are assumed to be parallel.

Geometrically, this can be understood as placing the simulation box at an effective comoving distance $D \gg \tilde{x}_{12}$, 
where $\tilde{x}_{12}$ is the mean separation between galaxies or halos. While at $z = 0.3$ and $z = 0.5$ the distant 
observer approximation is fully applicable, at $z = 0$ it should be interpreted as a redshift-space analysis without including projection effects.
To obtain redshift-space positions, we modify the object positions along the LOS according to:
\begin{equation}
    \vec{x}_{\mathrm{RSD}} = \vec{x} + \frac{1+z_{\mathrm{snap}}}{H(z_{\mathrm{snap}})} \left( \vec{v} \cdot \hat{e}_{\parallel} \right) \hat{e}_{\parallel},
\end{equation}
where $\vec{x}_{\mathrm{RSD}}$ and $\vec{x}$ are the redshift-space distorted and original comoving positions, respectively, 
$\vec{v}$ is the peculiar velocity, $\hat{e}_{\parallel}$ is the normalized LOS direction, and $H(z_{\mathrm{snap}})$ is 
the Hubble parameter at the snapshot redshift $z_{\mathrm{snap}}$.

To maximize the information extracted from the 3D velocity field, we distort the simulation box separately along each of the three 
Cartesian directions $[x, y, z]$, assuming the LOS along each axis in turn. Computations are performed independently for each LOS orientation. 
Although the large-scale three-dimensional peculiar velocity field is correlated, meaning that the three projections are not fully 
independent, this procedure effectively yields approximately $\sqrt{3}$ times more independent redshift-space realizations compared to a single projection.

\section{Averaged correlation functions and moments} \label{sec.calc}
We work with the three-dimensional volume-averaged correlation functions \cite{gaztanaga1998the,pollo1997gravitational} defined as
\begin{equation}
    \bar{\xi}_J(R) \equiv \langle \delta_R^J \rangle,
    \label{eq.XiDef}
\end{equation}
where $R$ is the smoothing scale.\footnote{For transparency, throughout this paper we denote the smoothing scale as $R$ 
regardless of whether it refers to real or redshift space.}
Here, $\delta_R = \rho_R / \bar{\rho}_R - 1$ denotes the density contrast, defined as the excess of the local density $\rho_R$ over 
the mean density $\bar{\rho}_R$.
The lower index subscript $R$ indicates that the density field is smoothed using a three-dimensional top-hat spherical window function 
of a co-moving radius $R$, and the angle brackets $\langle \cdot \rangle$ denote ensemble averaging over the entire catalog. 

When working with discrete data, the smoothed density $\rho_R$ at a given radius $R$ can be approximated by counting the number 
of objects within a sphere of radius $R$. As long as the number of objects $N \gg 1$, the discrete distribution provides 
a good approximation to a continuous background fluid \cite{peebles1980the,bernardeau2002large}.
We estimate $\bar{\xi}_J(R)$ by performing counts-in-cells (CIC) over random positions within the catalog and computing 
the central moments of the counts:
\begin{equation}
    M_J(R) = \mathrm{E}\left[ (N - \mathrm{E}[N])^J \right],
\end{equation}
where $\mathrm{E}[N]$ denotes the expected value of the counts $N$, after applying normalization and shot-noise corrections (see e.g. \cite{gaztanaga1994high}). 
Now, to get an estimate of $\bar{\xi}_J$ from Eqn. (\ref{eq.XiDef}) we just need to dive the central moments by $E[N]^J$.
In addition, we consider the reduced cumulants $s_J$ defined as
\begin{equation}
    s_J \equiv \frac{\bar{\xi}_J}{\bar{\xi}_2^{J-1}}.
\end{equation}

Under the assumption of a power-law power spectrum, perturbation theory predicts the reduced cumulants to be weakly varying 
monotonic functions of scale \citep{Bernardeau1994,bernardeau2002large}, making them useful diagnostics for cosmological tests.

The connection between volume-averaged correlations and the central moments shows that $\bar{\xi}_J(R)$ depends on the shape 
of the probability distribution function (PDF) of the counts. More precisely, these correlation functions are the connected parts of the moments, hence 
they quantify the departure from a Gaussian distribution. Simultaneously, the deviations of 
the PDF from an initially Gaussian form are driven by gravitational evolution of the density field (see \cite{pollo1997gravitational,amendola2010dark,schneider2015extragalactic}).

Thus, $\bar{\xi}_J(R)$ capture the growth of non-Gaussian features and quantify the gravitational structure formation 
and evolution of the large-scale Universe.

\section{Real and Redshift Space Skewness}\label{sec.res}

\begin{figure}[h!]
\centering
\includegraphics[width=0.9\columnwidth]{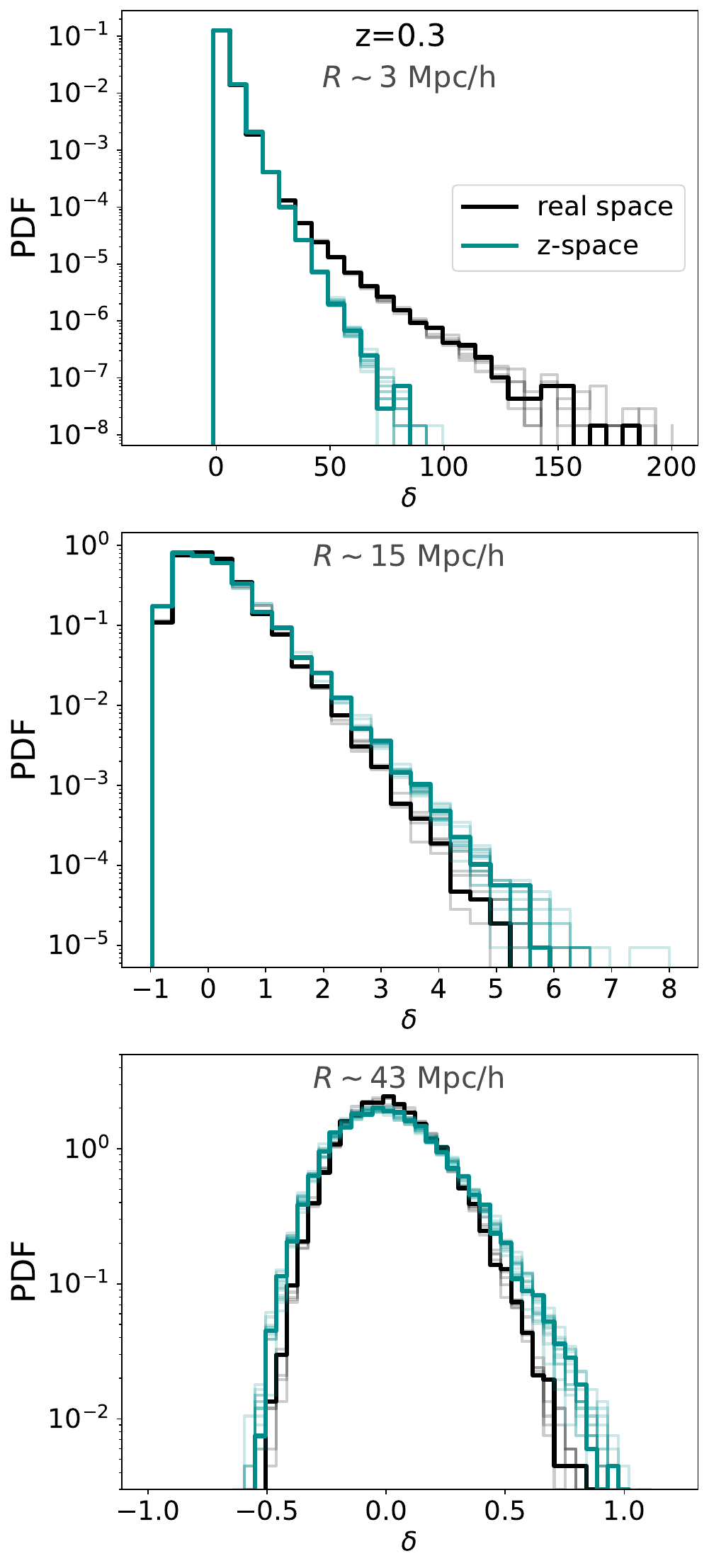}
\caption{Distribution of the density contrast of dark matter simulation pseudo-particles in real space (black) and redshift space (dark cyan) at $z=0.3$ for three chosen smoothing scales. Thinner semi-transparent lines correspond to PDFs from individual realizations, while the thick solid line shows the average over all realizations, separately for each scale and space. Note the different horizontal and vertical axis ranges.}
\label{fig.DM_counts}
\end{figure}

We begin our analysis by examining the statistical properties of the density contrast. Figure~\ref{fig.DM_counts} shows the probability distribution functions (PDFs) of the density contrast $\delta_R$, estimated from CIC by treating the discrete particle counts as a proxy for the local density. We display results at three chosen smoothing scales, both in configuration (real) space and redshift space, for the GR model at redshift $z=0.3$.

The histograms of counts converted into density contrast reveal clear RSD effects. At small scales ($R = 3\,h^{-1}\mathrm{Mpc}$), the density contrast is suppressed in redshift space due to the Fingers-of-God (FoG) effect \cite{kaiser1987clustering}. This is strikingly evident by comparing the long exponential tail of the real-space histogram, extending up to $\delta \sim 200$, with the much more compact redshift-space distribution, where few cells reach even moderate contrasts of $\delta \sim 100$. 

At intermediate scales ($R = 15\,h^{-1}\mathrm{Mpc}$), which mark the transition from the non-linear to the linear regime in classical gravitational instability theory, the real and redshift-space PDFs are remarkably similar. 

Finally, in the large cell volume limit ($R = 43\,h^{-1}\mathrm{Mpc}$), we observe that the density contrast PDF in redshift space is wider than that of real space. This is a clear manifestation of the so-called Kaiser squashing effect, where coherent large-scale infall motions enhance the redshift-space density around large-scale structures \cite{kaiser1987clustering}.

\begin{figure}[t]
\centering
\includegraphics[width=\columnwidth]{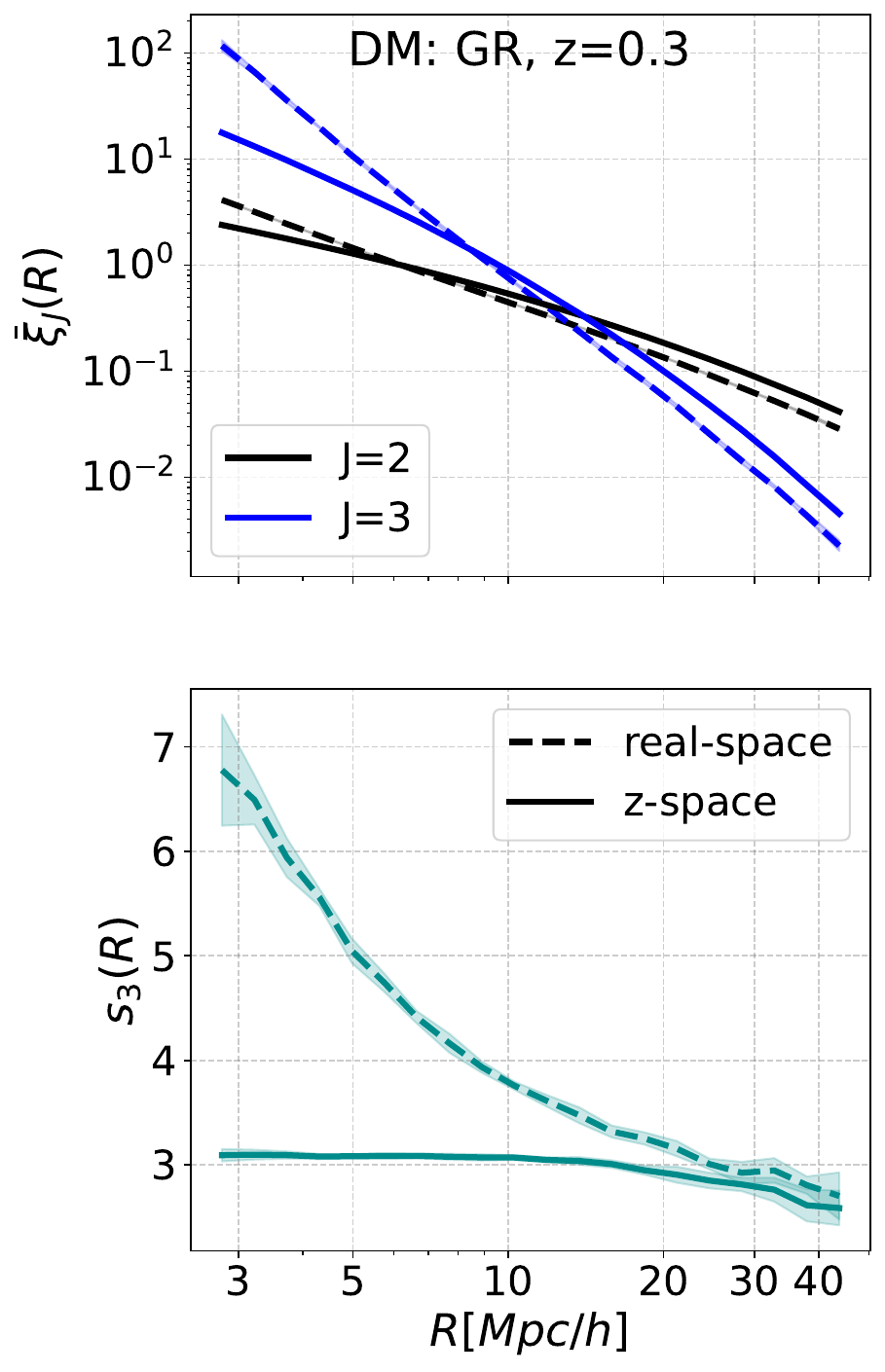}
\caption{The second- and third-order volume-averaged correlation functions (top panel) and the third-order reduced cumulant (bottom panel) at redshift $z=0.3$. Dashed lines correspond to real space, while solid lines to redshift space. All results are for the GR scenario for dark matter particles.}
\label{fig.DM_compare_GR}
\end{figure}

We can take a more quantitative view of the CIC distributions by examining the first two moments and the related $\bar{\xi}_J$ 
and $s_J$ statistics. Figure~\ref{fig.DM_compare_GR} shows the second- and third-order volume-averaged correlation functions (left panel) 
and the corresponding reduced skewness $s_3$ (right panel) for both real and redshift space at $z = 0.3$. We limit our analysis to scales 
$R \lesssim 40\,h^{-1}\mathrm{Mpc}$. For larger cell radii, the third moment of the CIC becomes significantly affected 
by finite-volume effects stemming from the combination of the simulation box size and the limited number of realizations \cite{baugh1995a}.

\begin{figure*}[t]
\centering
\includegraphics[width=0.9\textwidth]{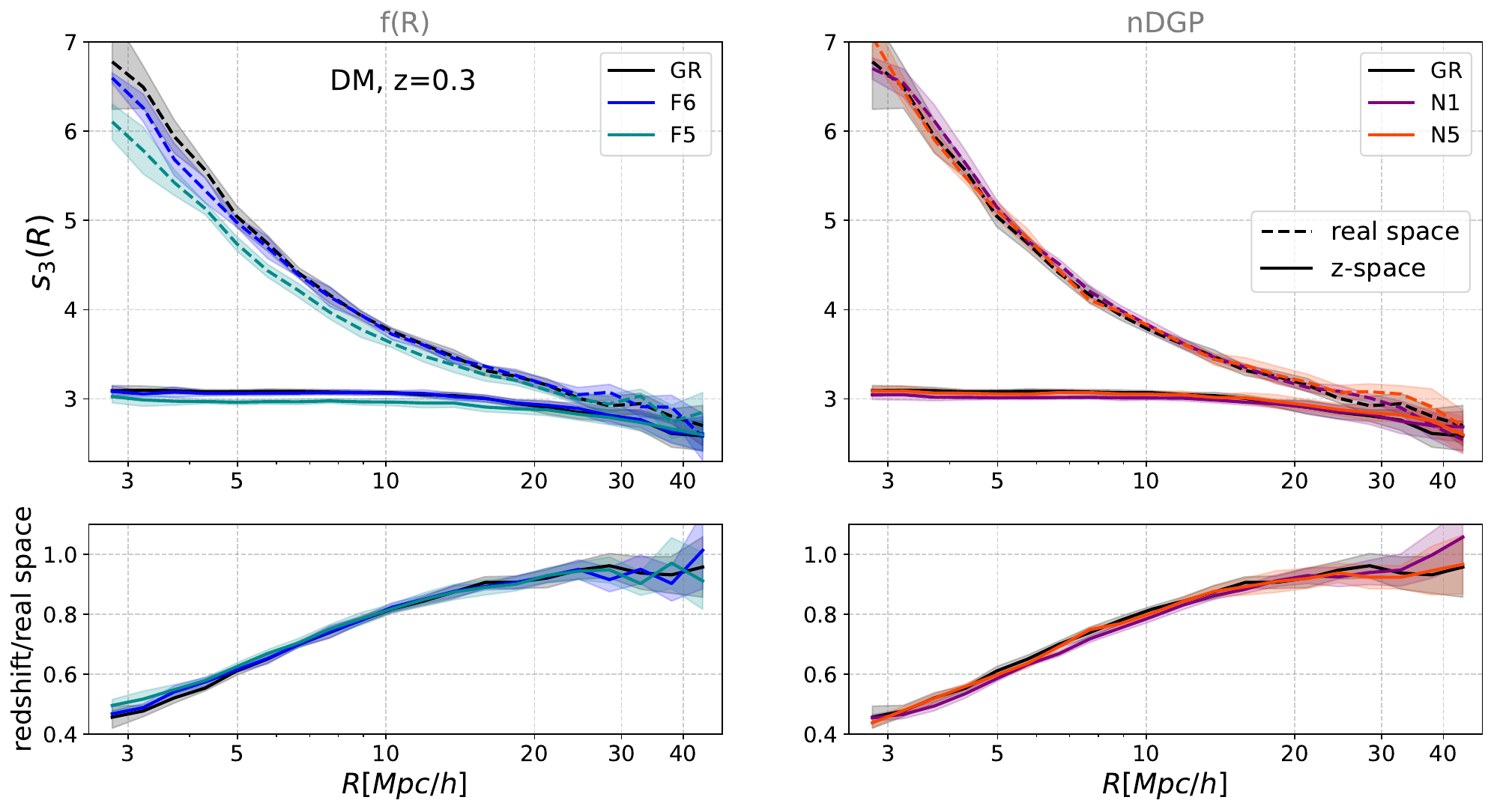}
\caption{Upper panels: third-order reduced cumulant $s_3$ of dark matter for GR (black), compared to $f(R)$ (left) and nDGP (right) gravity models at $z=0.3$. Dashed lines represent real space, and solid lines indicate redshift space. Lower panels show the ratio of $s_3$ between redshift and real space for the particular gravity models. }
\label{fig.DM_S3}
\end{figure*}

\begin{figure*}[t]
\centering
\includegraphics[width=0.9\textwidth]{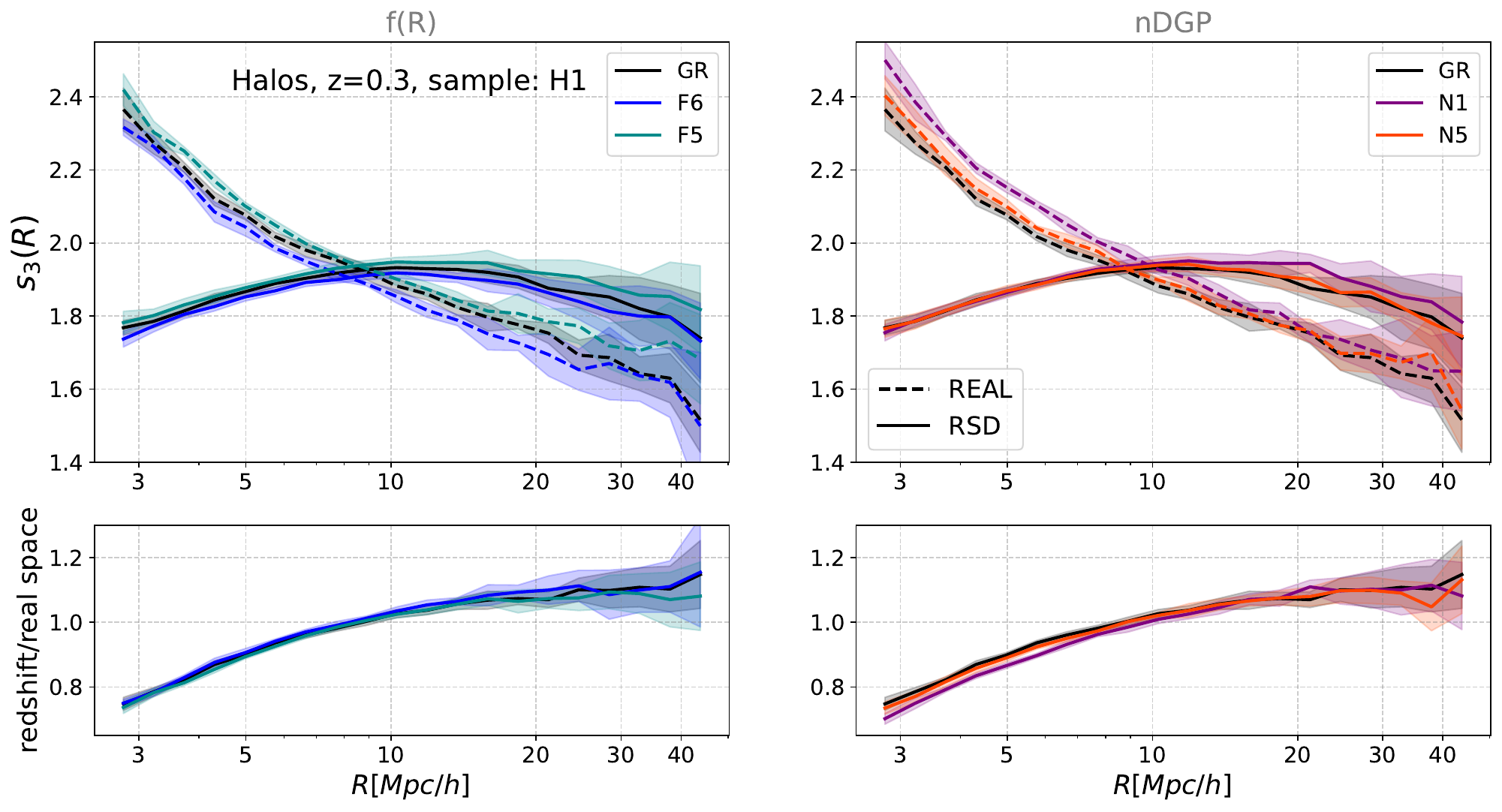}
\caption{Analogous plot as Fig. \ref{fig.DM_S3}, but for the halos from sample H1.}
\label{fig.halos_H1_snap31}
\end{figure*}

The most striking feature is the significant reduction of $s_3$ in redshift space for scales $R \lesssim 30\,h^{-1}\mathrm{Mpc}$ 
compared to real space. This suppression is caused by statistically random peculiar motions of objects on small scales, leading to 
the FoG effect.
This behavior of skewness was already reported by \cite{hivon1995redshift}, who showed that the redshift-space $s_3$ is much less 
scale-dependent  than its real-space counterpart. A similar suppression is also visible in the volume-averaged correlation functions 
$\bar{\xi}_J$ for $J = \{2,3\}$, manifesting as a reduction of the signal at $R < 10\,h^{-1}\mathrm{Mpc}$ in redshift space compared 
to real space. These trends are consistent with the previous results obtained from the CIC analysis (Figure~\ref{fig.DM_counts}).

At larger scales, redshift space yields slightly higher $\bar{\xi}_J$ values compared to real space. However, due to the construction 
of the cumulants, the overall trend 
of the $s_3$ amplitude reduction in z-space is preserved. The interpretation of
behavior of volume-averaged correlations is more convoluted, 
whereas cumulants show the real-to-redshift space transition that makes a more direct link to the predictions of 
perturbation theory (PT). Therefore, in this work, we focus primarily on the $s_J$ statistics, especially the skewness $s_3$.

Next, we compare different gravity models in the context of real and redshift-space. Figure~\ref{fig.DM_S3} shows 
the skewness $s_3$ for the GR, $f(R)$, and nDGP models. The trend of redshift-space suppression of the $s_3$ signal is quantitatively 
similar across all models. Even for the $F5$ model, which exhibits a significantly lower $s_3$ than the other models in real space, 
this difference relative to GR persists in redshift space.

On the other hand, the ratio of redshift- to real-space skewness, shown in the bottom panels, is in strong agreement among the different gravity scenarios.
We discuss the potential universality of the real-to-redshift-space transition in Section~\ref{sec.universalratio}. Higher-order
reduced cumulants $s_J$ with $J > 3$ exhibit similar behavior and we do not show them here.

As we checked, the skewness in redshift space also shows slight variations across redshifts. At $z = 0$, we observe that $s_3$ slowly decreases 
(by $\lesssim 10\%$) with scale between $3$ and $30\,h^{-1}\mathrm{Mpc}$, a feature that is not present at higher redshifts. 

\subsection{Dark matter halos}
\label{subsec:halos}

We now move from the counts of dark matter field tracers (particles) to the clustering analysis of halos. While the CIC 
of dark matter particles capture the properties of the smooth dark matter background field, it is the distribution of halos 
that is more directly related to observations. Halos are not only biased tracers of the underlying dark matter distribution 
but also host the luminous counterparts — galaxies. Thus, the analysis of the CIC distributions of the real and redshift-space 
halo fields provides information that has both stronger physical context and a more direct connection to observations.

The large-scale dark matter density field approximately follows a log-normal distribution, with positive skewness reflecting 
the continuous accumulation of matter into dense regions. Gravitational instability leads to the formation of collapsed halos, 
whose abundance — the halo mass function — depends primarily on the halo virial mass \cite{press1974formation,sheth2001ellipsoidal}.
Motivated by this, we divide our halo population into three samples based on different virial mass cuts (as listed in Table~\ref{tab.halomasscuts}) and analyze their hierarchical clustering separately.

We begin by examining the skewness of the H1 halo sample at $z = 0.3$, shown in Figure~\ref{fig.halos_H1_snap31}. This can be directly compared to the corresponding statistics for the smooth dark matter field presented in Figure~\ref{fig.DM_S3}.  
The first striking difference lies in the magnitude and scale dependence of RSD. For the smooth dark matter 
case, the $s_3$ amplitude was suppressed in redshift space up to $R \sim 30\text{--}40\,h^{-1}\mathrm{Mpc}$; for the H1 halo sample, however, we identify a qualitatively different redshift-space behavior with two distinct regimes. At small scales ($R \lesssim 10\,h^{-1}\mathrm{Mpc}$), 
the skewness is clearly suppressed in redshift space, similarly as was the case for the smooth density field. However, at larger separations 
($R \gtrsim 10\,h^{-1}\mathrm{Mpc}$), this trend is reversed: the amplitude of $s_3(R)$ is enhanced compared to real space. 
This behavior parallels the classical picture of the monopole of the two-point correlation function (2PCF): at small scales, 
random virial motions cause amplitude suppression (FoG effect), while at large scales, coherent infall enhances the clustering amplitude (Kaiser squashing effect \cite{kaiser1987clustering}). 

The Kaiser effect should be also present in the dark matter density field. For the case of our simulation scales and resolution
it appears, however, that it is subdominant even at larger scales, where the overall suppression from virial motions seems to
dominate the full $s_3$ amplitude. In contrast, for 
halos, where we only consider bulk motions of their centers of mass, the large-scale infall is dominating the velocity field, allowing the Kaiser enhancement to become significant at large-scales. Additionally, the overall amplitude of $s_3$ is consistently lower for halos than for the smooth 
dark matter field, reflecting the effect of halo biasing on hierarchical clustering \cite{Fry1993}.

From Figure~\ref{fig.halos_H1_snap31}, it is evident that the differences in skewness amplitude between GR and EG models identified in real space are strongly suppressed in redshift space. The fact that both the shapes and 
amplitudes of $s_3$ converge across GR and EG models in redshift space indicates that the overall effects of enhanced structure formation,
as fostered by EG, somehow conspire to bring the redshift-space skewness back close to GR values.  It seems that the more enhanced the real-space clustering becomes in a given EG model, the stronger the corresponding redshift-space suppression is, resulting in a near-cancellation of the two trends. 
This is particularly evident in the lower panels of Figure~\ref{fig.halos_H1_snap31}, where we show the ratio of skewness 
in redshift space to real space for all models.
Both $f(R)$ and nDGP models exhibit ratios very close to the fiducial GR case. 
A weak exception appears for the N1 model, where H1 halos exhibit slightly stronger redshift-space suppression at $R \lesssim 10\,h^{-1}\mathrm{Mpc}$ compared to GR. However, given the size of the $1\sigma$ scatter, this effect is small and likely 
difficult to be detected observationally.

Interestingly, if this near-universality of the real-to-redshift-space skewness transition also holds for other halo samples and 
for galaxies, this implies that knowledge of the real-space clustering alone could be sufficient to predict the skewness amplitude 
in redshift space, regardless of the underlying gravity model. Qualitatively similar trends are found for higher-order reduced 
cumulants such as kurtosis ($J = 4$) and hyper-kurtosis ($J = 5$) but we do not display them here, focusing just on skewness.

\begin{figure}[t]
\centering
\includegraphics[width=0.85\columnwidth]{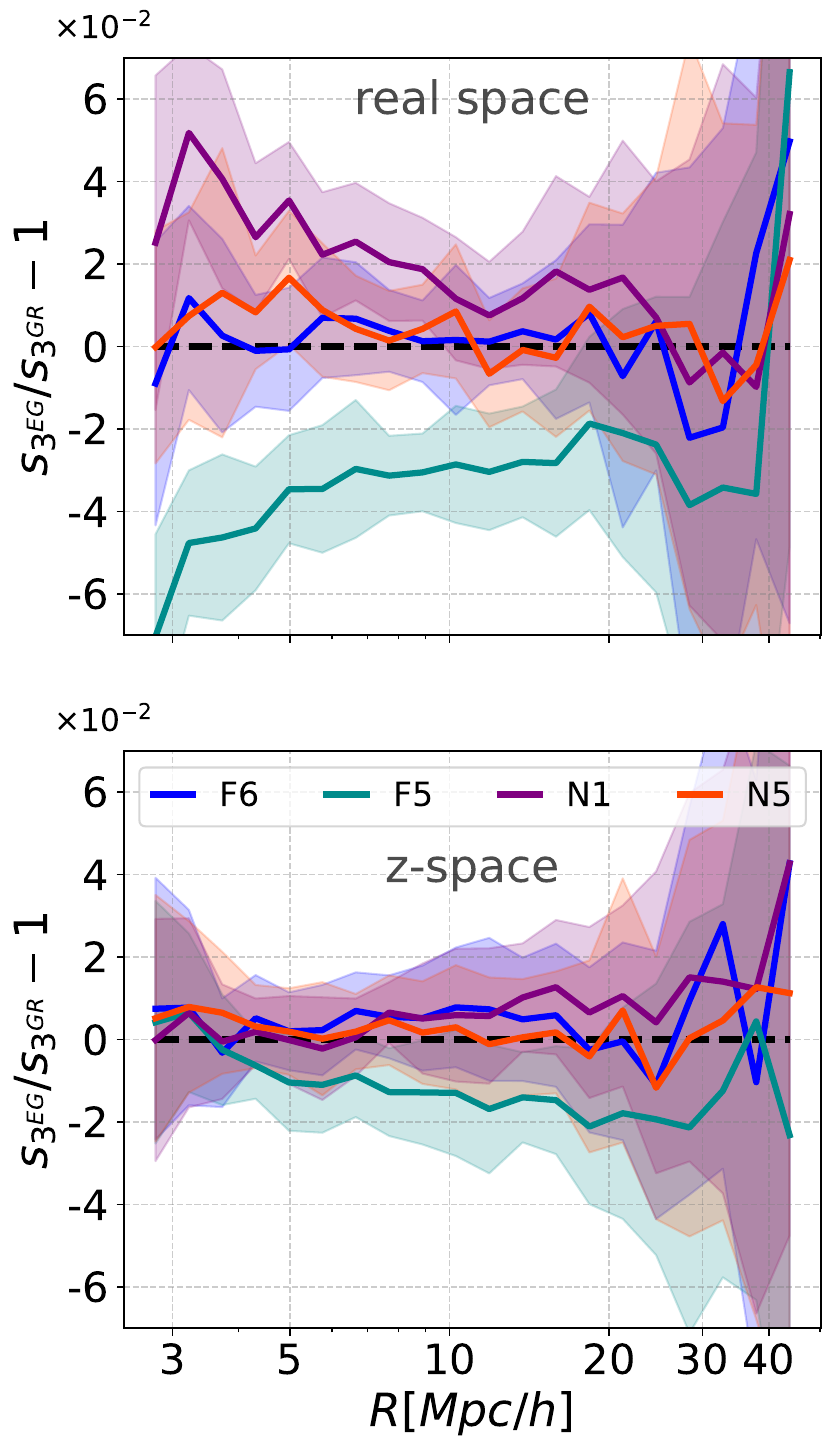}
\caption{Relative differences in the reduced third-order cumulant $s_3$ between EG models and GR for the halo sample H2 at $z=0.5$
for real (top panel) and redshift-space (bottom panel).}
\label{fig.halos_MGsignal_H2_0.5}
\end{figure}

So far, we have compared the skewness amplitudes between real and redshift space by considering all models together. We now 
investigate the actual relative differences between the EG models and the fiducial GR case taken as a reference. 
These results are shown in Figure~\ref{fig.halos_MGsignal_H2_0.5}.
For this exercise, we selected the $z = 0.5$ and H2 halo sample, since for this case the deviations from GR appear to be maximized. 
Examining the top panel, we recover the well-known results for real-space clustering (see e.g., Fig. 11 in \cite{Hellwing2017}) for a comparable halo sample. Here, the differences between various EG models and GR are typically contained within 5\% at small scales 
($R \lesssim 10\,h^{-1}\mathrm{Mpc}$), while at larger scales the differences become insignificant.
The situation is somewhat reversed when we move to redshift space, as illustrated in the bottom panel of 
Figure~\ref{fig.halos_MGsignal_H2_0.5}. At small scales ($R \lesssim 10\,h^{-1}\mathrm{Mpc}$), we observe convergence 
of $s_3$ for all models, with no significant deviations. Only at larger scales do some EG models show hints of departure 
from GR. The most notable case is the F5 model, which around $R \sim 20\text{--}30\,h^{-1}\mathrm{Mpc}$ exhibits $\sim2\%$ 
difference from GR, marginally reaching $1\sigma$ significance.

While considering different halo samples at $z = 0.5$, we found that a statistically significant signal for the F5 model is also 
present for the H3 sample, both in real and redshift space (at the $\sim 5\%$ level), whereas for the H1 sample, the signal
vanishes completely in redshift space.
Interestingly, at $z = 0.3$, the only significant deviation from the GR case is observed for the F5 model and 
the H3 halo sample. 
The H1 sample, which has the highest spatial abundance, would in principle be expected to provide the strongest signal-to-noise
ratio for CIC distribution functions. However, the suppression of the signal in redshift space persists.

\begin{figure}[t!]
\centering
\includegraphics[width=0.9\columnwidth]{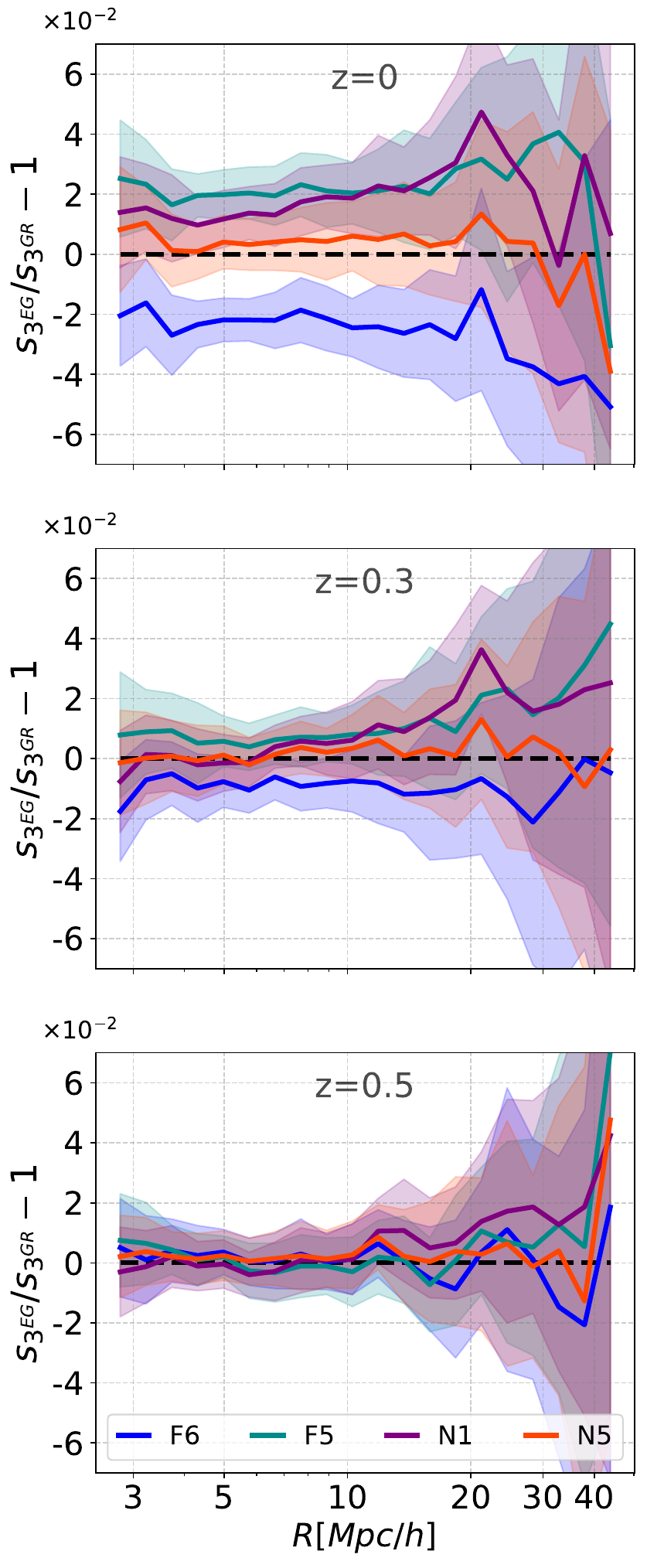}
\caption{Relative redshift-space differences in the reduced third-order cumulant $s_3$ between EG models and GR for the H1 halo sample at three redshifts: $z = 0$ (top), $z = 0.3$ (middle), and $z = 0.5$ (bottom).}
\label{fig.halos_3z_H1}
\end{figure}

For a more complete view, we present the results for the H1 sample collected across all three redshift epochs considered, as shown 
in Figure~\ref{fig.halos_3z_H1}.
Firstly, one can notice that the net differences between EG models and GR decrease with increasing redshift 
across all models considered. 
Interestingly, for this halo sample (i.e., H1), it is the F6 model that exhibits the strongest overall deviations from GR. 
Although initially surprising, this result is consistent with findings in the literature (see e.g., \cite{Shi2015,gupta2022universality}), 
which indicate strong non-linear behavior of F6 halos in the small-mass regime. As in the previous case of the smooth density field, we find qualitatively similar trends when examining higher-order moments, such as kurtosis and hyper-kurtosis.
The main takeaway from our halo sample CIC analysis is that, although significant differences in real-space 
skewness amplitudes exist between EG and GR cases, most — if not all — of the signal is suppressed when moving to counts performed in redshift space. This is a quite remarkable and important negative result. In the subsection below, we investigate how the situation changes 
when analyzing CIC for the mock galaxy sample.

\subsection{Mock Galaxies}
We now move to the observable tracers of the cosmic density field — galaxies. In this case, we work with a number density 
of $\sim 3.2 \times 10^{-4}\,(h/\mathrm{Mpc})^3$, which places our mock galaxy sample between the H2 and H3 halo datasets in terms 
of spatial abundance. We performed an analogous analysis to that presented above for halos.

We find that the nDGP models mostly match GR across all cases, except at $z = 0$ in real space, where $1\sigma$ deviations 
appear for the N1 model at some limited range of scales. However, these deviations disappear after moving to redshift space. For the $f(R)$ 
models, the situation is different. The F5 variant exhibits significant deviations from GR at every redshift in real space, 
and the signal partially survives in redshift space at $z = 0.5$ and, to a lesser extent, at $z = 0$ (at approximately 
the $1\sigma$ level). At $z = 0.3$, the F5 model shows strong deviations from GR, but only for the full
galaxy sample in 
real space. For the F6 model, $1\sigma$ deviations are observed that appear to be largely independent of redshift.

\begin{figure}[t!]
\centering
\includegraphics[width=0.85\columnwidth]{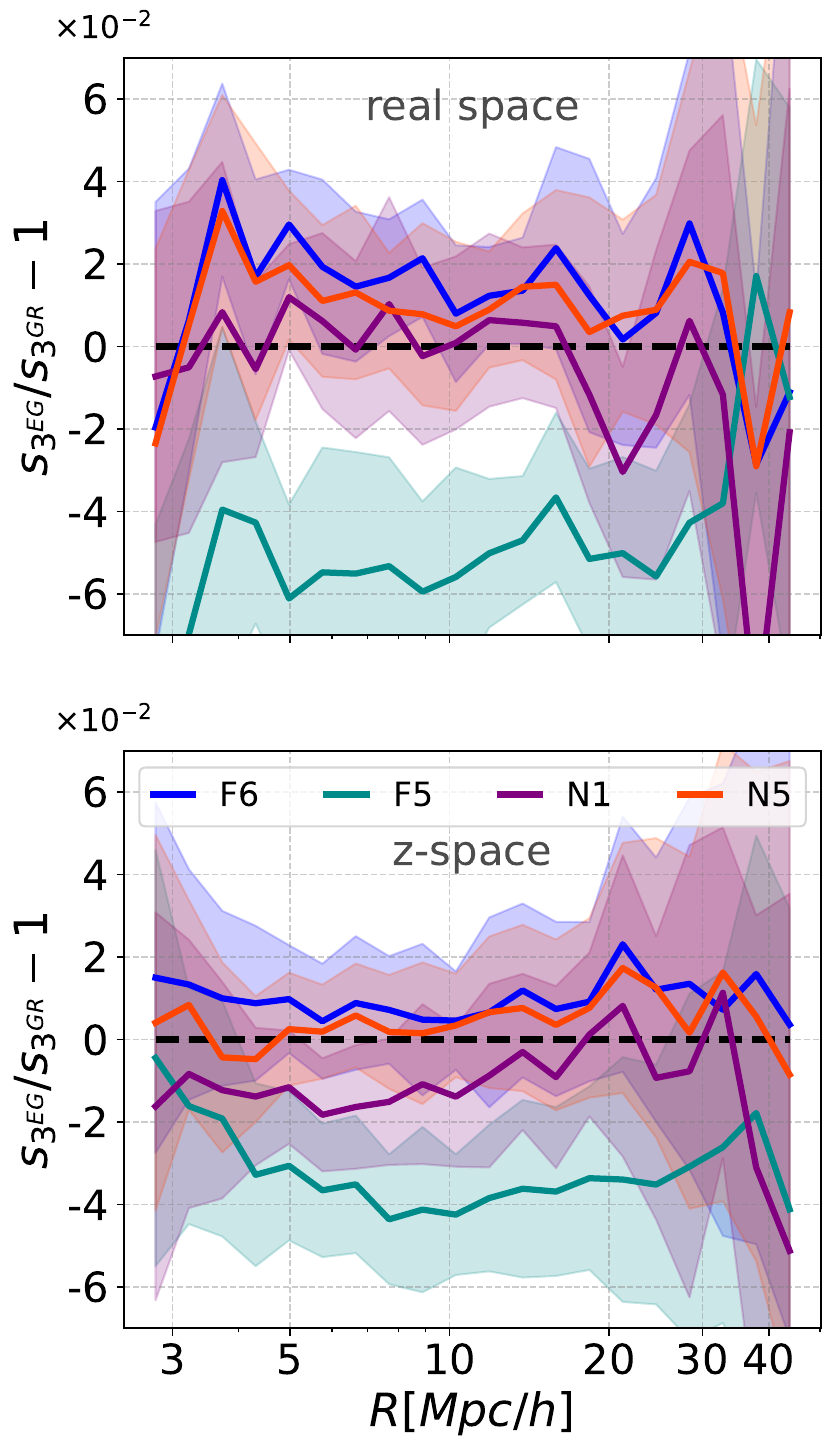}
\caption{Relative differences in the reduced third-order cumulant $s_3$ between EG models and GR for the full galaxy sample, comparing real (top) 
and redshift space (bottom) at $z = 0.5$.}
\label{fig.galx_snap27}
\end{figure}

Since signals in redshift space are the most relevant in the observational context, we focus our attention on the 
models at $z = 0.5$. Figure~\ref{fig.galx_snap27} shows the relative difference between EG and GR for the reduced skewness $s_3$, 
comparing the  models in both real and redshift space at this redshift.
One can notice a significant EG signal for the F5 model. Unlike F6, which drops below $1\sigma$ significance 
in redshift space, the F5 model maintains a $\sim4\%$ deviation with $\sim2\sigma$ significance in redshift space.
This result is in agreement with Refs. \cite{Hellwing2020the}.
For F6, 
the differences in redshift space are practically undetectable. Moreover, galaxies may in some cases provide a stronger EG signal 
in redshift space than halos. A comparison between Figure~\ref{fig.galx_snap27} and Figure~\ref{fig.halos_MGsignal_H2_0.5} shows 
that the deviation of F5 from GR is approximately twice as large for galaxies compared to the H2 halo sample, which provides the most 
comparable number density (see Table~\ref{tab.halomasscuts}). However, at lower redshifts, this enhancement is much less pronounced. 
This result is encouraging, as galaxies are the only directly observable large-scale tracers of the cosmic web. Although 
the $\sim 5\%$ difference observed for the H3 halo sample at $z = 0.5$ is slightly stronger than that for galaxies, the galaxy 
signal remains promising for observational constraints.

The signal of F5 shown in Figure~\ref{fig.galx_snap27} is particularly promising in this context. Results from \cite{wolk2013evolution} 
indicate uncertainties on skewness measurements of $\sim 5\text{--}15\%$ for selected galaxy samples at comparable redshifts
(see Fig. 12 therein). Given that those measurements were based on a survey covering approximately 155 square degrees,  
surveys with larger coverage should provide much tighter constraints on skewness.  Confronting such observational data with 
$s_3$ predictions for GR and F5 from mock galaxy catalogs mimicking the observations could potentially verify the validity of the models with high significance.

We find qualitatively similar trends and signals for higher-order reduced cumulants $s_{4,5}$
for the mock galaxies. In fact, the differences between GR and extended gravity models grow larger for $J = 4$ and $J = 5$; however,  
the associated uncertainties also increase more rapidly reducing the overall significance. Therefore, we focus our analysis on $s_3$.
\begin{figure}[t!]
\centering
\includegraphics[width=0.9\columnwidth]{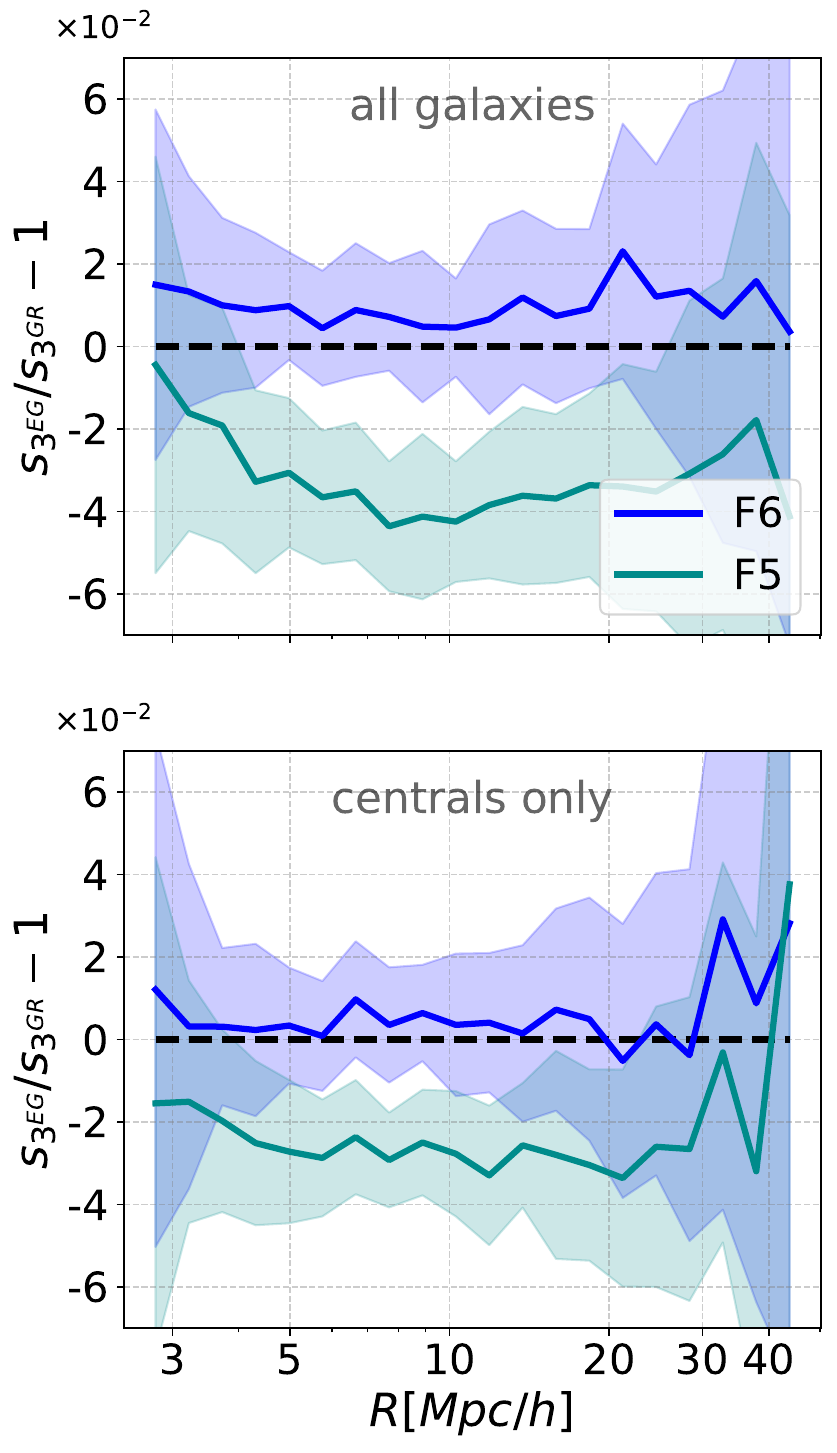}
\caption{Relative redshift-space differences in the reduced third-order cumulant $s_3$ between EG models and GR for mock 
galaxies, comparing the full sample (top) and centrals-only subsample (bottom), both at $z = 0.5$.}
\label{fig.galx_RSDratio_snap27}
\end{figure}

Up to this point, all results discussed refer to the full galaxy sample including both central and satellite galaxies. 
A notably different behavior of the skewness signal arises when considering only central galaxies in redshift space. 

Figure~\ref{fig.galx_RSDratio_snap27} shows that the skewness signal for the full galaxy sample is stronger than that for 
the centrals-only at most scales. 
At first glance, this behavior might seem surprising, since satellite galaxies, which typically reside within 
the virial radius of their host halos, are expected to contribute additional signal primarily at scales smaller than 
the average halo size, roughly $\lesssim 2\,h^{-1}\mathrm{Mpc}$ \cite{zheng2005theoretical,busha2011statistics,alam2021towards}. 

The strengthening of deviations from GR for the full sample is likely related to the enhanced small-scale nonlinearities 
introduced by the satellite population, particularly within the one-halo regime.\footnote{In $f(R)$ gravity, satellites 
do not experience fundamentally different forces compared to centrals on small scales, due to efficient screening within 
dense halo environments. However, their presence enhances the small-scale clustering signal, particularly within 
the one-halo regime, and thus indirectly contributes to the observed skewness deviations.}

The observed all-scale increase of the F5 deviation in the full sample is a manifestation of the nature of the statistics used.
Although the effects of satellites are most prominent at small distances, we use volume-averaged correlation functions in this work. Consequently, 
larger scales also incorporate contributions from smaller scales, allowing the satellite effect to propagate beyond the one-halo 
regime. We find that this effect persists at all redshifts considered, i.e $z = \{0, 0.3, 0.5\}$.

Another interesting feature we find is that for central-only galaxies, $s_3$ is smaller in real space than in redshift space at all scales, which is very distinct behavior than for both DM and halos. 
This inversion is indeed observed consistently across all gravity models we consider and contrasts with the behavior found for dark matter and halos, where real-space skewness typically exceeds redshift-space skewness.
This can be interpreted in two ways: either adding RSD in centrals-only galaxies increases the nonlinearities, or ignoring satellite galaxies in real space decreases the skewness at all scales, but primarily around the one-halo regime, whereas in redshift space the effect is distributed more uniformly across scales.
The first interpretation highlights that the centrals-only sample is much more selective compared to the full galaxy sample, leading to weaker spatial correlations on scales below the mean inter-galaxy separation. This is primarily due to the halo exclusion effect: in a centrals-only sample, no two galaxies can occupy the same halo, which suppresses the one-halo contribution to clustering. In contrast, including satellites reintroduces close galaxy pairs, reconstructing the missing small-scale clustering power associated with the one-halo term.
The velocity field encodes additional clustering information — for example, from satellites that are absent in the centrals-only sample. Therefore, the redshift-space the amplitude is boosted relative to real space.

In contrast, for the full galaxy sample, small-scale motions primarily act to decrease the strong skewness signal originating 
from satellites, which are now included in the catalog. We find that the aforementioned effects are present also at other redshifts, 
except for $R < 6\,h^{-1}\mathrm{Mpc}$ at $z = 0$ for the centrals-only sample, where the real-space skewness becomes 
approximately $10\%$ larger than in redshift space for the F5 model.

The second interpretation is more straightforward. Ignoring satellite galaxies in real space lowers the skewness primarily at 
the one-halo scale where satellites reside. However, due to volume averaging, this suppression also partially influences larger scales. 
In redshift space, much of the small-scale signal from satellites is suppressed by RSD. Hence, removing satellites does not 
induce as strong a scale-dependent change. Nonetheless, the additional nonlinearities encoded within the velocity field still enhance 
the skewness in the centrals-only redshift-space sample.

\subsection{Universality of Real to Redshift-Space Transition}\label{sec.universalratio}

Our results, both for the dark matter density field and for halo clustering, indicate that 
the ratio between skewness in redshift and real space exhibits a very close agreement across all EG models and the fiducial GR case. 
This finding provides a promising opportunity for predicting higher-order redshift-space clustering cumulants in EG scenarios 
based solely on real-space predictions. We discuss this feature further in the Appendix, where we additionally compare one 
example of the $s_3$ 
ratios with the ratios of the volume-averaged correlation functions $\bar{\xi}_2$ and $\bar{\xi}_3$, from which the skewness $s_3$
is constructed.
While these results hold robustly for the dark matter and halo populations the situation is different for galaxies. 
Here, the agreement between redshift-to-real-space skewness ratios across different gravity models 
is limited to specific cases and redshifts, largely due to the probabilistic nature of the halo occupation distribution (HOD) method used to construct 
the mock catalogs. In particular, the random assignment of satellite galaxies introduces additional small-scale peculiar 
velocities that are not fully correlated with the underlying matter density field. This synthetic velocity component partially 
decouples the galaxy clustering from the local density environment, making it more difficult to preserve the universality of 
the redshift-to-real-space ratio observed for dark matter haloes \citep[see \eg][]{Tinker2006, Hikage2013}.

\section{Summary and Conclusions}\label{sec.concl}
In this work, we have investigated the skewness of the three-dimensional dark matter and galaxy density fields across different 
redshifts and gravity scenarios, focusing on the impact of redshift-space distortions. The comparison between real and redshift 
space revealed several remarkable trends, which we summarize below:
\begin{itemize}
    \item For dark matter density, the skewness $s_3$ is suppressed in redshift space relative to real space at all scales and for all gravity models (Figure~\ref{fig.DM_S3}).    
    \item For halos, skewness is not universally suppressed in redshift space (Figure~\ref{fig.halos_H1_snap31}). At large scales, $s_3$ is enhanced due to the Kaiser squashing effect, which is damped in dark matter particles by random motions.    
    \item Differences between GR and EG models are suppressed in redshift space compared to real space, especially for halos. This significantly limits the utility of skewness as a gravity diagnostic for $f(R)$ and nDGP models.    
    \item Up to $\sim5\%$ deviations between GR and EG are found for halos in real space (Figure~\ref{fig.halos_MGsignal_H2_0.5}), but redshift-space differences are generally smaller ($\sim2\%$) and only marginally significant at $\sim 1\sigma$.    
    \item For the H1 halo sample (number density of $10^{-3}$ $(\mathrm{Mpc}/h)^{-3}$), deviations in EG from GR decrease with redshift. Interestingly, the F6 model shows the strongest deviations for H1, whereas F5 dominates for the sparser H2 and H3 samples (Figure~\ref{fig.halos_3z_H1}).    
    \item For galaxies, nDGP models do not show significant deviations from GR in redshift space (Figure~\ref{fig.galx_snap27}). In contrast, the F5 $f(R)$ model maintains a $\sim 2\text{--}4\%$ deviation.    
    \item Galaxies in redshift space can reveal stronger EG signals than halos. Comparing Figure~\ref{fig.galx_snap27} and Figure~\ref{fig.halos_MGsignal_H2_0.5} shows that the F5 deviation is about twice as large for galaxies than for halos with similar number densities.    
    \item Separating out central-only galaxies from the full sample
    reveals additional features (Figure~\ref{fig.galx_RSDratio_snap27}). Samples containing both centrals and satellites show stronger skewness signals for F5, although for the F6 model the difference is negligible.    
    \item The ratio of skewness between redshift and real space appears nearly universal across gravity models, especially for $f(R)$ (Figure~\ref{fig.halos_H1_snap31}). This glimpse of universality is discussed in the Appendix, where we find that while $s_3$ ratios match well across models, the underlying $\bar{\xi}_2$ and $\bar{\xi}_3$ ratios differ.
\end{itemize}

Our analysis demonstrates significant and complex effects of redshift-space distortions on higher-order clustering statistics. 
For the dark matter density field, skewness is universally suppressed in redshift space due to virial small-scale motions 
generating the Fingers-of-God effect. Halo populations exhibit a more nuanced behavior, with suppressed skewness at small 
scales transitioning into enhanced skewness at intermediate scales ($\sim 10\,h^{-1}\mathrm{Mpc}$) due to coherent large-scale 
infall motions (Kaiser squashing).

Critically, the differences in skewness between GR and EG scenarios, prominent in real space, become much 
less pronounced in redshift space. Nevertheless, the F5 $f(R)$ model maintains a statistically significant deviation from GR at 
approximately $4\%$ level for galaxy samples, suggesting potential observational detectability, particularly with 
the current new stage and large-scale galaxy surveys. 

Interestingly, we uncover that the ratio of skewness between redshift and real spaces is remarkably stable across different gravity
models. This approximate universality implies that real-space clustering predictions can be effectively used to estimate 
redshift-space skewness, greatly simplifying theoretical modeling for EG scenarios.

Furthermore, our study reveals enhanced sensitivity of galaxy skewness compared to halo skewness in redshift space, 
especially when satellite galaxies are included. Satellites thus provide critical clustering information, reinforcing their 
importance in cosmological analyses aiming to test gravity models.

In summary, our findings highlight both challenges and opportunities in using higher-order clustering statistics to test gravity.
While redshift-space distortions complicate direct interpretations of deviations from GR, our results point to viable pathways 
for utilizing observational data, emphasizing the role of satellite galaxies and the robustness of skewness ratios as powerful diagnostic tools.

\section{Acknowledgments}
The authors would like to thank Enrique Gaztañaga for valuable and engaging discussions at the various stages of this project.
This work is supported via the research project ‘COLAB’ funded by the National Science Center, Poland, under
agreement number UMO-2020/39/B/ST9/03494

\appendix

\section{Real to Redshift-Space Transition}
\label{sec:appendix}

In this section, we discuss the ratio between statistics computed in redshift and real space. We define the ratio as:
\begin{equation}
    \eta_J = \frac{X_{J,\mathrm{RSD}}}{X_{J,\mathrm{REAL}}},
\end{equation}
where $X_J$ can refer either to the volume-averaged correlation functions $\bar{\xi}_J$ or to the reduced cumulants $s_J$, 
depending on whether ratios of correlations or cumulants are being considered. The subscripts RSD and REAL denote redshift and 
real space, respectively.

We find that the ratios for cumulants are, in many cases, universal with respect to the underlying gravity model. In this Appendix, 
we focus on halo clustering, as constructing halo catalogs comparable to ours requires relatively few additional assumptions. 
The situation for mock galaxy samples is more complicated due to the necessity of modeling specific selection functions, stellar masses, 
and luminosities. At the same time, the redshift-space clustering of the smooth dark matter density field cannot be directly observed 
or easily connected to measurable quantities. Thus, halos offer the most practical case for investigating potential universal relations 
between configuration and redshift space.
We find that in many cases, the ratio $\eta_J$ agrees between gravity models, particularly for $s_J$, but not necessarily 
for $\bar{\xi}_J$. Here, the nDGP model family however generally perform much worse than $f(R)$, indicating much lower degree of universality.

\begin{figure}[t]
\centering
\includegraphics[width=\columnwidth]{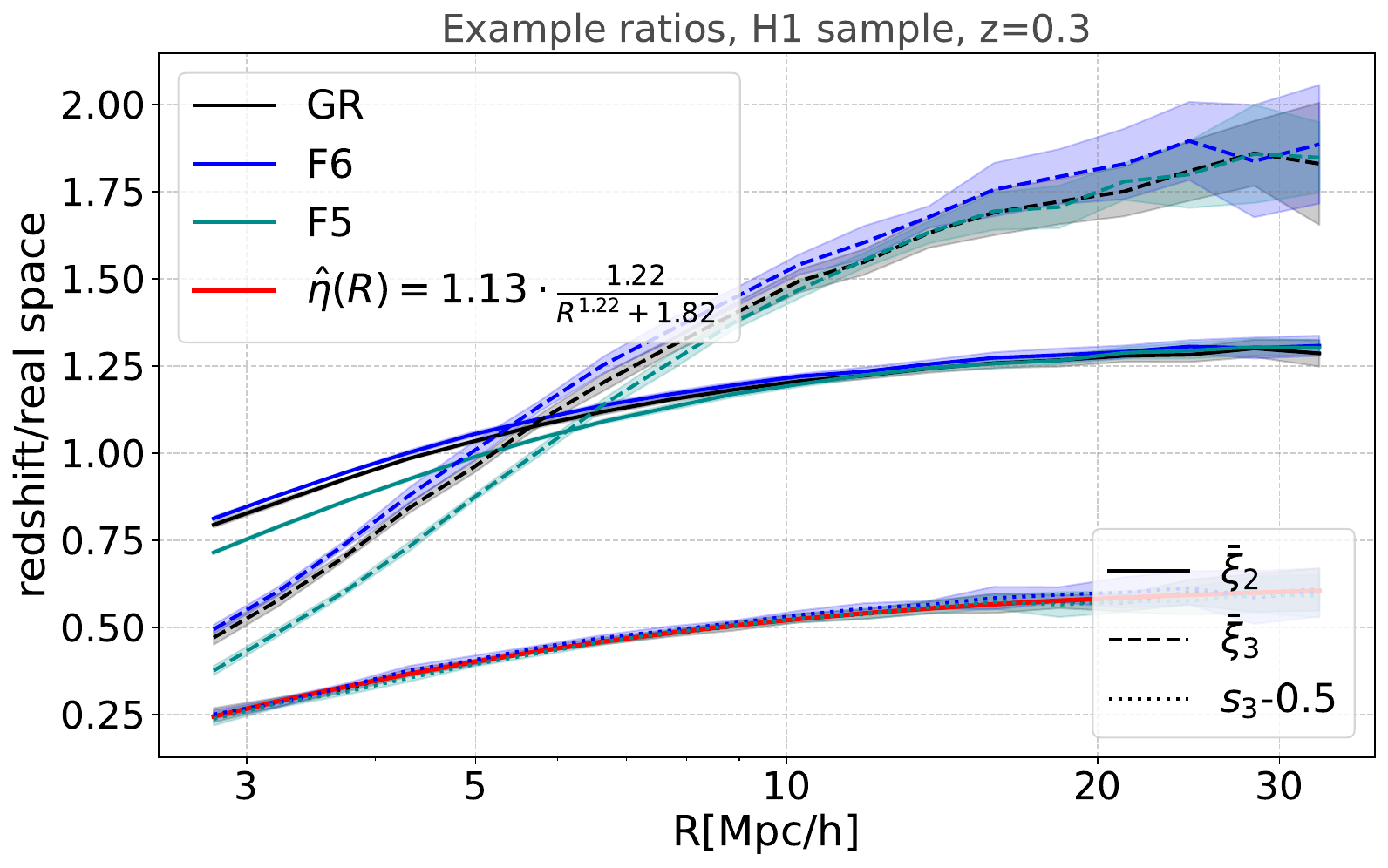}
\caption{Real-to-redshift space ratios at $z = 0.3$ for the H1 halo sample. The comparison is made in terms of $\bar{\xi}_2$ (solid lines), $\bar{\xi}_3$ (dashed), and $s_3$ (dotted) for the $f(R)$ gravity scenarios considered. The empirical fit for the $s_3$ case (Eq.~\ref{eq.fit}) is marked in red.
For clarity, the $s_3$ results have been offset downward by 0.5.}
\label{fig.ratios}
\end{figure}

Figure~\ref{fig.ratios} shows an example for the H1 halo sample at $z = 0.3$, for $f(R)$ and GR.
For this example, we additionally fit the ratio $\eta_J(R)$ with an empirical function:
\begin{equation}
    \hat{\eta}_J(R) = A \times \frac{R^{\alpha}}{R^{\alpha} + B},
    \label{eq.fit}
\end{equation}
where $R$ is the smoothing scale in $h^{-1}\mathrm{Mpc}$, and $A$, $\alpha$, and $B$ are free parameters.
Due to finite-volume effects — i.e., cosmic variance limited by the simulation box size — the statistical uncertainties grow at 
larger scales. To ensure statistically robust fits, we restrict our analysis to scales $R < 50\,h^{-1}\mathrm{Mpc}$. 
The ratios for other halo samples, redshifts, and higher orders are also well fitted by a function of the form given in Eq.~\ref{eq.fit}. 
However, we do not provide the fitting parameters for all cases, as they strongly depend on the specific simulation details and 
do not carry physical meaning. Instead, we demonstrate the potential universality of $\eta_J$ and provide an empirical fitting form that 
could be tested in future works.

The main findings regarding the real-to-redshift space transition can be summarized as follows:
\begin{itemize}
    \item Differences in the $\eta_J$ ratios between $f(R)$ models and GR are generally strongly suppressed when considering cumulants. 
    We observe this effect consistently across all samples and redshifts.
    \item In most cases, one can reliably estimate the redshift-space skewness for both GR and extended gravity scenarios based on 
    the universality of $s_3$.
\end{itemize}

The finding that the redshift-to-real-space ratio does not depend significantly on the assumed gravity model is promising for recovering 
real-space skewness from redshift-space observations. Furthermore, this effect can be tested against results from other simulations, 
helping to better constrain background fifth-force screening mechanisms in redshift space.

\bibliography{main}

\end{document}